\documentclass[aps,prd,preprint,superscriptaddress,showpacs]{revtex4}
\usepackage{graphicx}
\usepackage{verbatim}
\usepackage{ulem}
\usepackage{color}
\definecolor{My_red}        {cmyk}{0.00,1.00,1.00,0.20}

\newcommand{\bmat}{\left(\begin{array}}
\newcommand{\emat}{\end{array}\right)}
\newcommand{\beq}{\begin{equation}}
\newcommand{\eeq}{\end{equation}}
\newcommand{\wt}{\widetilde}

\def\ra{\rightarrow}

\def\Ld{\Lambda}
\def\ld{\lambda}
\def\f{\frac}

\def\bwt{\begin{widetext}}
\def\ewt{\end{widetext}}
\def\be{\begin{equation}}
\def\ee{\end{equation}}
\def\bea{\begin{align}}
\def\eea{\end{align}}
\def\bean{\begin{align*}}
\def\eean{\end{align*}}
\def\bary{\begin{array}}
\def\eary{\end{array}}
\def\bit{\begin{itemize}}
\def\eit{\end{itemize}}

\def\ra{\rightarrow}

\def\Ld{\Lambda}

\def\ld{\lambda}

\def\su5u1{SU(5) \times U(1)}
\def\fsu5u1{SU(5) \times U(1)_X }
\def\so10{SO(10)}
\def\sq20{SO(10) \times SO(10)}

\def\ra{\rightarrow}

\def\Ld{\Lambda}
\def\ld{\lambda}
\def\f{\frac}

\def\L{\left(}
\def\R{\right)}
\def\bwt{\begin{widetext}}
\def\ewt{\end{widetext}}
\def\be{\begin{equation}}
\def\ee{\end{equation}}
\def\bea{\begin{align}}
\def\eea{\end{align}}
\def\bean{\begin{align*}}
\def\eean{\end{align*}}
\def\bary{\begin{array}}
\def\eary{\end{array}}
\def\bit{\begin{itemize}}
\def\eit{\end{itemize}}

\def\ra{\rightarrow}

\def\Ld{\Lambda}

\def\ld{\lambda}

\def\su5u1{SU(5) \times U(1)}
\def\fsu5u1{SU(5) \times U(1)_X }
\def\so10{SO(10)}
\def\sq20{SO(10) \times SO(10)}

\usepackage[centertags]{amsmath}
\usepackage{amssymb}

\begin{document}

\title{Origins of the Isospin Violation of Dark Matter Interactions}

\author{Xin Gao}

\affiliation{Key Laboratory of Frontiers in Theoretical Physics,
      Institute of Theoretical Physics, Chinese Academy of Sciences,
Beijing 100190, P. R. China }

\author{Zhaofeng Kang}

\affiliation{Key Laboratory of Frontiers in Theoretical Physics,
      Institute of Theoretical Physics, Chinese Academy of Sciences,
Beijing 100190, P. R. China }

\author{Tianjun Li}

\affiliation{Key Laboratory of Frontiers in Theoretical Physics,
      Institute of Theoretical Physics, Chinese Academy of Sciences,
Beijing 100190, P. R. China }

\affiliation{George P. and Cynthia W. Mitchell Institute for
Fundamental Physics, Texas A$\&$M University, College Station, TX
77843, USA }

\date{\today}

\begin{abstract}

Light dark matter (DM) with a large DM-nucleon spin-independent cross section
and furthermore proper isospin violation (ISV) $f_n/f_p\approx-0.7$ may provide a way to understand
the confusing DM direct detection results. Combing with the stringent astrophysical and collider constraints, we
systematically investigate the origin of ISV first via general operator analyses
and further via specifying three kinds of
(single) mediators: A light $Z'$ from chiral $U(1)_X$, an approximate spectator
Higgs doublet (It can explain the $W+jj$ anomaly simultaneously) and color triplets. In addition,
although $Z'$ from an exotic $U(1)_X$ mixing with $U(1)_Y$ generating
$f_n=0$, we can combine it with the conventional Higgs to achieve proper ISV.
As a concrete example, we propose the $U(1)_X$ model where the $U(1)_X$ charged light sneutrino is the inelastic DM,
which dominantly annihilates to light dark states such as
$Z'$ with sub-GeV mass. This model can address the recent GoGeNT annual modulation consistent with
other DM direct detection results and free of exclusions.

\end{abstract}



\maketitle

\section{Introduction and motivation}

The way that the dark matter (DM) interacts with visible matters is a puzzle due to the
absence of confirmative experimental results. The recent possible progresses made on direct
detections may shed light on it. On the one hand, the DAMA/LIBRA~\cite{DAMA} and CoGeNT~\cite{CoGeNT}
experiments report events, which point to a light DM (LDM) having mass $\sim$8 GeV and a rather
large spin-independent (SI) recoil cross section with nucleons, $\sigma_{n}^{\rm SI}\sim10^{-4}$ pb (The supscript
SI will be implied). On the other hand, the XENON~\cite{XENON} and CDMS~\cite{CDMSII}
experiments report null results, which challenges the CoGeNT/DAMA results. Moreover,
for the CoGeNT/DAMA favored DM there may be some expected but not found signals
from astrophysics or colliders. So how to reconcile these results guides us to identify some of the DM properties.

DM-nucleon interactions with isospin violation (ISV) may provide a way to reconcile
various direct detection results~\cite{ISV}. DM ISV is not a
novel phenomena~\cite{ISVO,Kang:2010mh}, and it arises when the DM-proton and
DM-neutron interactions have different strengths, namely $f_{p}\neq f_n$.
In particular, if $f_n/f_p<0$, the DM-proton and DM-neutron scattering amplitudes will
destructively interfere, leading to a cancellation in the DM-nucleus scattering amplitude.
The degree of cancelation varies with the target nucleus used in different experiments.
Ref~\cite{ISV} showed that if $f_n/f_p=-0.7$, we can not only substantially weaken the XENON100 constraint
but also make the CoGeNT- and DAMA-region overlap. This scenario has tension with the CDMS-Ge experiment
which uses the same nucleus as
the CoGeNT experiment. But if we only consider the annual modulation results (The observed CoGeNT annual
modulation has significance of 2.8$\sigma$~\cite{CoGeNTA}), the inelastic DM (iDM)
scenario~\cite{iDM} is able to enhance the modulation and thus reduce the tension
between the CDMS-Ge and CoGeNT experiments. Ref.~\cite{Frandsen} found that by taking
$f_n/f_p=-0.7$ and the quenching factor $Q_{Na}$ close to its upper-limit $0.43$,
one can address all the confusing experimental results via an iDM $\sim$10 GeV with a
mass splitting $\delta\simeq 15$ keV. Additionally, right now the direct detection
experiments may have reached the level to measure the individual strength of $f_n$ and
$f_p$~\cite{Pato:2011de}. In a word, it is worthy of studying the origin of DM ISV systematically.

In our analyses, we focus on the scalar and fermionic DMs, which are required
to present the following features:
\begin{itemize}
  \item A proper ISV $f_n/f_p\approx -0.7$. How to get this ISV is highly non-trivial.
  We will show that conventional mediators like the Higgs boson, $Z$ boson and squarks
   fail to accommodate this value, at least for single mediator case. Thus
 we need to investigate new mediators beyond them.
  \item A large DM-nucleon SI scattering cross section $\sigma_{n}\sim  10^{-2}$ pb.
  In the ISV scenario, the nucleus amplitude is reduced by the destructive interference,
  so $\sigma_{n}$ is required to about 2 orders larger than the conventional scenario.
  That large $\sigma_n$ will bring tensions with some astrophysical
  or collider constraints.
 \item The right DM relic density $\Omega_{\rm DM}h^2\simeq0.11$.
  We allow DM annihilation channels whose final states are not the
 standard model (SM) light fermions. Such channels actually appear in the
 complete models where LDM can annihilate into light dark sector states. This point
 will be very helpful somewhere.
\item If possible, the DM models should have connection with other new physics.
 Especially, we try to account for the recent Tavetron CDF W+2jets anomaly~\cite{Wjj}.
\end{itemize}

The paper is organized as follows. In Section~II, we make
some general  analysis on the dark matter with ISV based on effective operators.
In Section~III,  we turn to
their possible simple UV origin by specifying the mediators. Next, a concrete model with sneutrino iDM is presented.
 Conclusion and discussions are given in Section~\ref{CD}, and finally some
useful formulas are collected in the Appendices.

\section{Generic Operator Analyses}

In this section we will investigate the ISV origin based on the effective operator analyses.
The DM candidate is assumed to be either a scalar $\phi$ or fermion $\chi$. Relevant constraints are also collected to
constrain the DM interactions.

\subsection{Effective operators for CoGeNT/DAMA in the ISV scenario}\label{GOA}

The generic effective description on the interactions between DM and the SM fields involves
a large class of operators, e.g., ${\cal O}_{\rm DM}\bar f\Gamma f$ where $\Gamma=1,\gamma_\mu,\gamma_5...$ and
$f$ denotes the SM fermion and ${\cal O}_{\rm DM}$ denotes the DM bilinear such as $\bar \chi\chi,\bar \chi\gamma^\mu\chi...$,
and so on. But they are greatly reduced if we just interest in the operators which are relevant to the
CoGeNT/DAMA experiments.

In the first place, we pick out the operators which can generate DM-nucleon SI scattering cross sections.
Such operators are limited. Furthermore, if the corresponding cross sections
are not non-relativistic (NR) suppressed, the SM degrees of
freedom in these operators must constitute one of the following three operators~\cite{Agrawal:2010fh,Fan:2010gt}:
 \begin{align}
\bar q q,\quad \bar q \gamma_\mu q,\quad
(G_{\mu\nu}^a)^2.
\end{align}
Obviously, the gluon operator can not generate ISV and then are dropped in what follows. Recovering
the DM-bilinears, we then obtain the operators of interest:
 \begin{align}\label{DDS}
&a_{q_i}\bar \chi\chi \bar q_i q_i,\quad\quad\quad a_{q_i} \phi^\dagger\phi \bar q_i q_i,\\
&b_{q_i}\bar \chi\gamma^\mu\chi\bar q_i\gamma_\mu q_i,\quad b_{q_i}\phi^\dagger\tensor{\partial}^\mu\phi
 \bar q_i\gamma_\mu q_i.\label{DDV}
\end{align}
The operators in the first and second line lead to scalar and vector interactions between DM and nucleons, respectively.
$a_q$ and $b_q$ label the corresponding operator coefficients which are assumed to be suppressed
by the mass scale of some mediator $\Ld\gg M_{\rm DM}$, unless otherwise specified. Note that
if DM is a CP self-conjugate particle,
i.e., $\chi$ is a Majorana and $\phi$ is a real scalar, its vector interaction vanishes automatically.
In our notation, $q=u,d$ denote for the up- and down-type quarks respectively, and $i$ is the family index
(All fields are written in the mass eigenstates). We will show that only the first family quarks are crucial to
produce ISV. To suppress the potential large flavor violation, the quark bilinears are supposed to be
diagonal in the flavor space, but we will find that this assumption does not hold so naively.

There are some other operators which do not contribute to $\sigma_n$
in the NR limit, but they are usually generated together with Eq.~(\ref{DDS}) and Eq.~(\ref{DDV}) in a UV completed theory.
Moreover, these operators are relevant to determine the DM relic density and indirectly detect DM.
So we list them as the following:
 \begin{align}\label{ID}
&\bar \chi\gamma^5\chi\bar f\gamma^5 f,\quad \bar \chi\chi\bar f\gamma^5 f,\quad \bar \chi\gamma^5\chi\bar f f,\cr
&\bar \chi\gamma^5\gamma_\mu\chi\bar f\gamma^5\gamma^\mu f,\quad\bar \chi\gamma_\mu\chi\bar f\gamma^5\gamma^\mu f,\quad
 \bar \chi\gamma^5\gamma_\mu\chi\bar f\gamma^\mu f,\cr
 & \bar \chi\sigma^{\mu\nu}\chi\bar f\sigma_{\mu\nu} f,\quad  \bar \chi\sigma^{\mu\nu}\gamma^5\chi\bar f\sigma_{\mu\nu} f,
\cr
&|\phi|^2\bar f\gamma_5 f,\quad
\phi^\dagger\tensor{\partial}_\mu\phi\bar f\gamma^\mu\gamma^5 f,
\end{align}
where $f$ should be so light that DM can annihilate into it, with corresponding rate casted in Appendix~\ref{CS}.
Now Eqs.~(\ref{DDS})-(\ref{ID}) give a general effective description on the DM models inspired by the CoGeNT/DAMA experiments.
Having established this setup, we will investigate the ISV originating from the scalar and
vector interactions, respectively.

First let us consider scalar interactions shown in Eq.~(\ref{DDS}). Above all, we need to translate the ISV
of the microscopic DM-quark interaction into the ISV of the DM-nucleon interaction. This can be done in a conventional
way, i.e., constructing the effective theory describing the DM-nucleon interaction from the interaction at the quark level.
We have to figure out the quark bilinear matrix elements in the nucleon states which, in the case of scalar interaction,
are given by
 \begin{align}
m_q\langle n| \bar qq| n\rangle=m_n f_{T_q}^{(n)},
\end{align}
where $n$ denotes either the proton or neutron and $m_q \,(m_n)$ is the quark (nucleon) mass. $f_{T_q}^{(n)}$ are the form factors. For the
light quarks they take such values: $f_{T_u}^{(p)}=0.020\pm0.004$, $f_{T_u}^{(n)}=0.014\pm0.003$, $f_{T_d}^{(p)}=0.026\pm0.005$,
$f_{T_d}^{(n)}=0.036\pm0.008$, and $f_{T_s}^{(p,n)}=0.118\pm0.062$~\cite{NF}. As for the heavy quarks $q=c,b,t$, they contribute
to the nucleon mass through the triangle diagram~\cite{HQC}, and the corresponding nucleon matrix elements are given by
 \begin{align}
m_q \langle n|\bar qq| n\rangle=\f{2}{27}m_n\L 1-\sum_{q=u,d,s} f_{T_q}^{(n)}\R.
\end{align}
Thus the universal form factor (for heavy quarks) is
 \begin{align}
f_{T_G}^{(n)}=1-\sum_{q=u,d,s} f_{T_q}^{(n)},
\end{align}
which is $0.84$ and $0.83$ for the proton and neutron respectively.

Now the effective operators describing the DM-nucleon scalar interaction are $f_n\bar\chi\chi\bar nn$ and $f_n\phi^\dagger\phi\bar nn$
with the effective coupling
 \begin{align}\label{scalar}
f_n\propto{a_n}=&\,m_n\left[\sum_{q=u,d,s}a_q \f{f_{T_q}^{(n)}}{m_q}+\f{2}{27}f_{T_G}^{(n)}\sum_{c,b,t}\f{a_q}{m_q}\right]\equiv\sum_{q=\rm quarks}B_q^na_q,
\end{align}
where $f_n=a_n$ for a fermionic DM while $f_n=a_n/2M_{\rm DM}$ for a scalar DM, because in this case
$a_q$ has dimension $-1$ and moreover $\sigma_n$ is written in the form of Eq.~(\ref{DNS}).
Hereafter the $1/2M_{\rm DM}$ factor will be absorbed into $a_q$.
The dimensionless quantities $B_q^n\equiv f_{T_q}^{(n)}m_n/m_q$ are independent on DM interactions, encoding
the ISV in the nucleon itself. Using quark masses $m_u=0.002$ GeV, $m_d=0.005$ GeV, $m_s=0.095$ GeV, $m_c=1.25$ GeV,
$m_b=4.2$ GeV and $m_t=172.3$ GeV, then we get
\begin{align}\label{ISC}
&B_u^p\approx 9.3,\quad B_u^n\approx 6.5,\quad B_d^p\approx 5.1,\quad B_d^n\approx 7.1,\\
&B_s^{p,n}\approx 1.2,\quad B_c^{p,n}\approx 0.05,\quad B_b^{p,n}\approx 0.015,\quad B_t^{p,n}\approx 0.00035.\label{ISV}
\end{align}

Remarkably, from Eqs.~(\ref{scalar})-(\ref{ISV}) it is obvious that only the DM and $u/d$ interactions break isospin effectively.
Immediately, we draw the  conclusion: \emph{If scalar interactions account for the CoGeNT/DAMA experiments, the interactions between
DM and the first family quarks must give the predominant contribution}. Thus, at the nucleon level $f_n={\cal I}f_p$, where ${\cal I}\neq1$
measures the degree of ISV, means that at the quark level we should make $a_u$ and $a_d$ satisfy
 \begin{align}\label{ISVSquark}
\f{a_u}{a_d} ~\simeq~ \f{{\cal I}B_d^p-B_d^n}{B_u^n-{\cal I}B_u^p}.
\end{align}
The ratio is about $-0.77$ for ${\cal I}=-0.7$ (Throughout this work, we shall use it as a referred value of ISV and 8 GeV as the
referred DM mass). Furthermore, the effective DM-proton coupling can be organized in a form
 \begin{align}\label{ISVSfp}
f_p ~\simeq~\L\f{B_d^pB_u^n-B_d^nB_u^p}{B_u^n-B_u^p{\cal I}}\R\times a_d.
\end{align}
$f_p$ is factorized into the DM-quark effective coupling multiplying a model-independent factor casted in the bracket, which takes value
$-2.5$ for ${\cal I}\simeq-$0.7.

Next we turn to the vector interactions. The quark bilinear matrix elements in the nucleon states are greatly simplified by virtue of
the conservation of the vector current, to which the sea quarks and gluons do not contribute. As a consequence, the effective operators for
the DM and nucleons interactions are simply given by
 \begin{align}\label{vector}
 &{\cal L}_{vec}=b_n\bar\chi \gamma_\mu \chi \bar n\gamma^\mu n,\quad b_n\phi^\dagger \tensor\partial_\mu \phi \bar n\gamma^\mu n,\cr
 &f_p= b_p=2b_u+b_d,\quad f_n= b_n=2b_d+b_u.
\end{align}
Even in the case they are generated by integrating out a colored mediator instead of a heavy vector boson,
the above descriptions still hold. Then analogously to Eq.~(\ref{ISVSquark}), ISV at the nucleon level $b_n={\cal I}~b_p$ entails a ratio
\begin{align}
\f{b_u}{b_d}=\f{2-{\cal I}}{2{\cal I}-1}.
 \end{align}
It takes a value  $-9:8$ when ${\cal I}=-0.7$. Obviously, again \emph{the ISV from vector interactions must originate from the interactions
between the mediator and the first family quarks}. The DM-proton effective coupling in this case is expressed as
\begin{align}\label{fpV}
f_p=\f{3}{2{\cal I}-1}b_d,
 \end{align}
and $f_p=-1.25b_d$ for ${\cal I}=-0.7$.

Comments are in orders: (i) For the scalar interactions, the ISV in the DM-nucleon interaction
depends on both the ISV in the nucleon itself, i.e., $B_{u,d}^{(p)}\neq B_{u,d}^{(n)}$, and the ISV hidden in the DM-quark
interactions. While for the vector interactions, it is totally determined by the latter. (ii) In the ISV scenario we need
$\sigma_p\sim10^{-2}$ pb, so numerically we have the ratio $\langle\sigma_{an}v\rangle/\sigma_p\sim 10^2$ with
$\langle\sigma_{an}v\rangle\simeq1$ pb the required annihilation rate of thermal DM. On the other hand, a typical operator $\bar q q\bar\chi\chi$
($q$ denotes a light quark) gives the ratio
\begin{align}\label{estimation}
\f{\langle\sigma_{an}v\rangle}{\sigma_{p}}\sim \f{M_{\rm DM}^2}{\mu_p^2},\quad \mu_p\simeq 1\rm\,GeV.
\end{align}
Interestingly, for LDM around 10 GeV it is just at the aforementioned order. This \emph{numerical coincidence
involves three basic elements of DM, the mass, relic density and scattering rate}. Thus the
GoGeNT/DAMA inspired LDM models with ISV is ``justified" to some degree. Note that the estimation in Eq.~(\ref{estimation})
ignores velocity suppressing, but it still makes sense on account of the operator $(\bar\chi\gamma_5\chi)(\bar q\gamma_5q)$,
which has no velocity suppressing, usually does exist in a complete theory and moreover has comparable operator coefficient
with $(\bar\chi\chi)(\bar qq)$. Such arguments apply to other cases such as $\phi^\dagger\phi\bar qq$.

\subsection{Some constraints}\label{GOAC}

\begin{figure}[htb]
\begin{center}
\includegraphics[width=3.7in]{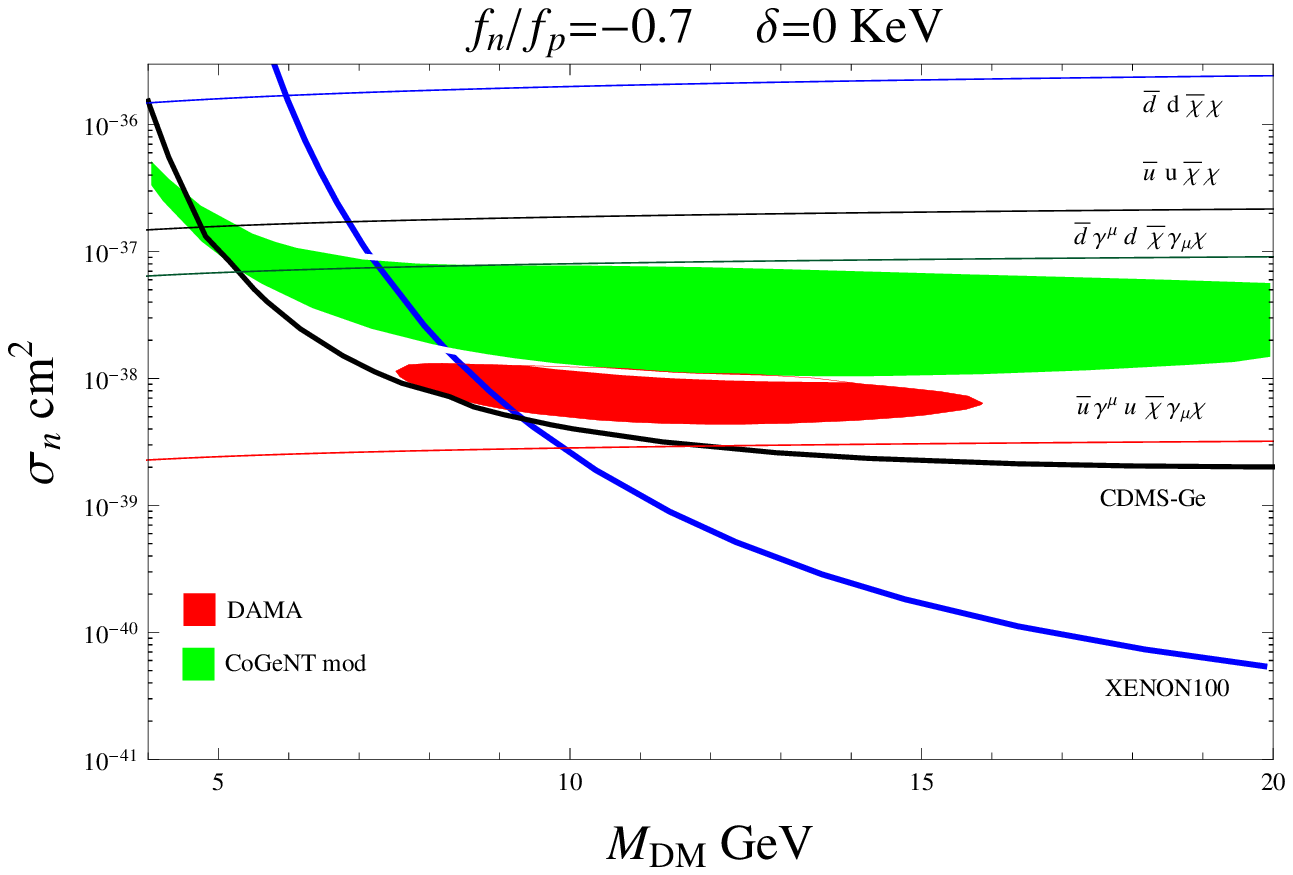}
\includegraphics[width=3.7in]{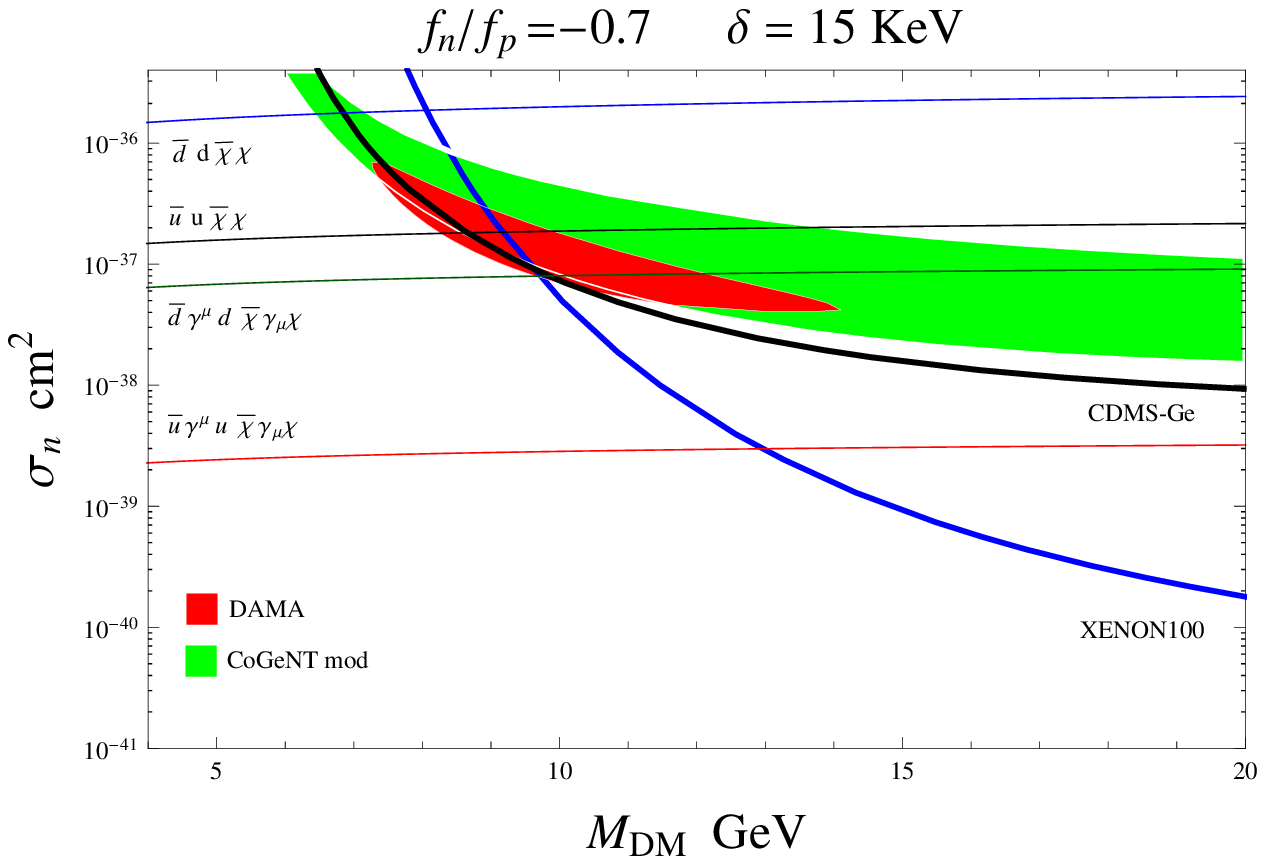}
\end{center}
\caption{\label{Fig1}Top: favored regions and exclusion contours in the ($M_{\rm DM},\sigma_n$) plane, ISV $f_n/f_p=-0.7,\delta=0$; Bottom (for the annual modulation):
ISV $f_n/f_p=-0.7,\delta=15$ keV.
 Collider exclusion for some operators are also imposed. The data are from Ref.~\cite{Frandsen} and Ref.~\cite{Y. Bai}.}
\end{figure}

The 8 GeV LDM with a quite large DM-nucleon scattering rate suffers a list of constraints, including the cosmological,
astrophysical and collider constraints. We denote them as ${\cal C}_{1-4}$ in the following:
\begin{description}
  \item[${\cal C}_1$: PAMELA]
  The PAMELA measures the antiproton spectrum from 1-100 GeV, which shows no deviation from the background~\cite{PAMELA}.
  On the other hand, due to the crossing symmetry, LDM may lead to the low-energy
  antiproton excess~\cite{PAEMALDM,Lavalle:2010yw}
  since it couples to quarks with significant strength. But the PAMELA constraint
  only applies to the DM annihilating into quarks with
  a rate larger than $0.1$ pb, so it is avoided if the rate has
  velocity suppressing (A preferred case in this paper). Moreover, the constraint
  can also be avoided by properly choosing astrophysical parameters~\cite{PAEMALDM}.
  \item[${\cal C}_2$: Solar Neutrino] Because of the large DM-nucleon cross section,
the solar captures DM particles with a large rate. DMs subsequently (cascade)
annihilate into neutrinos, which are expected to be observed by the Super-Kamiokande.
No observation of such signals excludes thermal DM with dominant annihilation modes into
$\tau\bar\tau/\nu\bar\nu/4\tau$ as well as the heavy quarks modes
$\bar b b/\bar c c$ ~\cite{Kappl:2011kz}. We have to
emphasize that this constraint is so strong that generically
one has to sufficiently suppress the DM annihilation rates through these channels, especially directly to neutrinos.
   \item[${\cal C}_3$: CMB]
   DM annihilations at redshifts $z\sim500- 1000$ may distort the Cosmic Microwave Background (CMB) spectrum,
   and the lack of such distortion gives a constraint on the GeV scale DM~\cite{CMBC}. To make DM being a thermal relic,
   its annihilation modes should be dominated by the $\mu$ or $\tau$ modes (and the corresponding 4 leptons), or
 the DM annihilation rate is velocity dependent.
        \item[${\cal C}_4$: Colliders] $\sigma_p$ can be converted to the
 DM production rate at the hadronic colliders like Tevatron, and the lack of relevant signals provides
    another constraint~\cite{collider,Y. Bai}. It is of great concern in the ISV scenario since it
    needs a large $\sigma_p$. The vector interactions accounting for $\sigma_p$
(even as weak as $0.001$ pb) have been definitely excluded, and the scalar interactions
are also on the brink of exclusion, as shown in Fig.~\ref{Fig1}. Consequently the scenario of iDM
with ISV is excluded. However, the collider constraint is draw in terms of effective operators with
a cut-off scale much higher than the DM mass. Therefore the constraint is invalid if the mediators,
which generate DM-quark interactions, have a mass $M_m$ much lighter than $M_{\rm DM}$.
In that case, $\sigma_p$ gets a great enhancement because $\sigma_n\propto 1/M_{m}^4$ while
the DM production rate does not since it scales as $1/s$. As a result the constraints
in~\cite{collider,Y. Bai} are evaded.

In addition, Ref.~\cite{Fox:2011fx} gives the LEP bound on the operators
${\cal O}_{\rm DM}\bar e \Gamma e$, and shows that this operator is not allowed to provide
the main annihilation channel for a thermal DM.
\end{description}

Let us summarize the constraints and point their possible implications (A recent more quantitative
study see Ref.~\cite{Mambrini:2011pw}). Among them, ${\cal C}_1$ can be satisfied and is thus not very sever.
While ${\cal C}_2$ is a rather stringent constraint even if DM annihilation is velocity dependent,
and it implies that \emph{thermal DM annihilates into $\mu$ or light quarks}
(The $e^\pm$ mode is excluded by the LEP in ${\cal C}_4$). In the ISV scenario, light quark modes are thus
favored, e.g., in the models given in Section~\ref{IHiggs} and Section~\ref{ColorTriplets}.
But ${\cal C}_3$ may disfavor it except that the annihilation rate is suppressed by velocity today
(Taking the astrophysical uncertainty into account, we may also regard ${\cal C}_3$ only as a
referred constraint). Last but not the least, ${\cal C}_4$ gives the most powerful constraint
in the actual model building. It picks out the models whose \emph{DM-quark interaction is either
a scalar type or mediated by a very light gauge boson}.

\section{ISV in the $SU(3)_C\times U(1)_{\rm EM}-$UV completions}

ISV at the effective operator level can not give more information, while figuring out the ISV origin
in the $SU(3)_C\times U(1)_{\rm EM}-$UV completed models provides more guidance on actual model building.
Thus in this section we will specify mediators which connect DM and the quarks, and four kinds of
mediator will be discussed.

\subsection{Chiral  $U(1)_X $ Vector Boson}\label{ISVU(1)}

A simple  way to produce ISV is to introduce an exotic $U(1)_X $ gauge boson
as the mediator. Before proceeding to a discussion on $Z'$,
we briefly prove that the $Z$ boson in the  SM can only generate
ISV with $|f_n/f_p|\ll 1$. Explicitly, the interactions between the $Z$ boson and
SM quark neutral current are given by
\begin{align}\label{}
{\cal L}_{NC}\supset \f{1}{\cos\theta_w}& \left[\bar u_L\gamma^\mu \L\f{1}{2}-\f{2}{3}\sin^2\theta_w\R u_L
+\bar u_R\gamma^\mu\L-\f{2}{3}\sin^2\theta_w\R u_R
    \right.& \nonumber \\
 &\left.
+\bar d_L\gamma^\mu \L-\f{1}{2}+\f{1}{3}\sin^2\theta_w\R d_L+\bar d_R\gamma^\mu \L+\f{1}{3}\sin^2\theta_w\R d_R\right] Z_\mu~.~\,
\end{align}
$\theta_w$ is the Weinberg angle. Utilizing the formula Eq.~(\ref{vector}), one can easily get
\begin{align}\label{}
\f{b_p}{b_n}=-\L1-4\sin^2\theta_w\R\approx -0.08\ll 1.
\end{align}
We emphasize that this result is completely determined by the SM structure.

Now consider  $Z'$ from an exotic $U(1)_X$. In terms of the discussions in the last Section, we only need to
consider a light $Z'$ (But the following discussions apply to any $Z'$ boson with
mass much larger than the typical transfer momentum scale). We will investigate what
kind of quark $U(1)_X$ charge assignment can generate the required ISV, and the implications to model building are
also discussed. For generality, we start from the $Z'$ and SM fermion current couplings
\begin{align}\label{}
{\cal L}_{NC}=-g_X \sum_{i,j}\bar f_i\gamma^\mu\left[(Q_{f_L})_{ij}P_L+(Q_{f_R})_{ij}P_R\right]f_jZ_\mu',
\end{align}
where $g_X $ is the gauge coupling of $U(1)_X $ and $Q_{f_{L/R}}$ are the
charge matrices of the left- and right-handed fermions in the family space.
To not induce large tree-level flavor changing neutral currents, it is reasonable to
assume $Q_{f_{L/R}}$ are diagonal matrices (it must be true for Abelian gauge group).
Focusing on the quark sector, we first transform quark flavors into the mass eigenstates
via $q_{L/R}\rightarrow V_{q_{L/R}}^\dagger q_{L/R}$ (The same letters are used
to label gauge basis and mass basis) and then get
\begin{align}\label{Z'S}
{\cal L}_{NC}'=&-g_X \sum_{q=u,d;i,j}\bar q_i\gamma^\mu\left[\L V_{q_L}Q_{{q_L}}V_{q_L}^\dagger\R_{ij} P_L+\L V_{q_R} Q_{{q_R}}V_{q_R}^\dagger\R_{ij} P_R\right]q_jZ'_{\mu}\cr
\supset&
-\f{g_X }{2}\sum_{q,i}\left[(Q_{q_L})_i|(V_{q_L})_{1i}|^2 +(Q_{q_R})_i|(V_{q_R})_{1i}|^2\right]\bar q\gamma^\mu  q Z'_{\mu}+...
\end{align}
In the second line, we only explicitly show terms involving the first family, while
other terms, involving the second and third families or axial vector currents, are denoted by the dots.

In this paper, the charge matrices are further assumed to be family universal $(Q_{q_{L/R}})_i\equiv Q_{q_{L/R}}$,
and thus due to the unitarity of $V_{q_{L/R}}$, the SM fermion mixings will not induce any flavor changing
neutral currents in Eq.~(\ref{Z'S}). In that case, the first-family quark charges account for the proper ISV:
\begin{align}\label{VISV}
\f{b_u}{b_d}=\f{Q_{u_L}+Q_{u_R}}{Q_{d_L}+Q_{d_R}}=\f{Q_{q_L}+Q_{u_R}}{Q_{q_L}+Q_{d_R}}.
\end{align}
In the second equation relations $Q_{u_L}$=$Q_{d_L}$$= Q_{q_L}$ have been used, because $u_L$ and
$d_{L}$ comes from the same $SU(2)_L$ multiplet $q_L$. Obviously, the ISV effect is
ascribed to the difference between the $u_R$ and $d_R$ charges. Immediately we draw such a conclusion:
Quarks must form chiral representations of $U(1)_X$. Therefore, we are forced to work in the two Higgs doublets
model (2HDM) where $H_{u}$ couples to up-type quarks and $H_d$ couples to
down-type quarks and charged leptons.

The number of free charges in Eq.~(\ref{VISV}) can be reduced further in a more realistic model.
First, the $U(1)_X $ gauge invariance of Yukawa interactions give charge equations
\begin{align}\label{}
Q_{q_L}-Q_{u_R}+Q_{H_u}=0,\quad Q_{q_L}-Q_{d_R}+Q_{H_d}=0~.~\,
\end{align}
Next, if we do not introduce additional colored fermions at low energy,
to cancel the mixed $SU(3)_C-SU(3)_C-U(1)_X$ anomaly (disregarding other anomalies),
quark charges need to satisfy the following condition,
\begin{align}
3\L Q_{q_L}-\f{Q_{u_R}}{2}-\f{Q_{d_R}}{2}\R=0~.~\label{SU(3)U}
\end{align}
From these equations, we get $Q_{H_u}=-Q_{H_d}$. Finally, combining them and Eq.~(\ref{vector}), we get
the charge conditions to obtain the proper ISV:
\begin{align}\label{solution}
Q_{u_R}=&\f{7-5{\cal I}}{6(1-{\cal I})}Q_{H_u},\quad
Q_{d_R}=-\f{5-7{\cal I}}{6(1-{\cal I})}Q_{H_u},\quad
Q_{q_L}=\f{1+{\cal I}}{6(1-{\cal I})}Q_{H_u}.
\end{align}
${\cal I}=-$0.7 gives rise to a somewhat peculiar solution: $Q_{u_R}=35/34Q_{H_u}$,
$Q_{d_R}=-33/34Q_{H_u}$ and $Q_{q_L}=1/34Q_{H_u}$~\footnote{A recent reanalysis given
in Ref.~\cite{Farina:2011pw} requires charge assignment ${\cal I}=1/5.4$, in turn
$Q_{u_R}=5/4Q_{H_u},Q_{d_R}=-3/4Q_{H_u},Q_{q_L}=1/4Q_{H_u}$, which is a natural pattern. Anyway,
the $U(1)_X$ charges are quite sensitivity to the actual data, and in this paper
we will not fix the concrete values.}.
However, if Eq.~(\ref{SU(3)U}) is relaxed by introducing new colored vector-like fermions,
the more elegant solution can be found. But such a possibility is beyond the scope of this work.
Additionally, it is not difficult to check that such $U(1)_X$ can not originate from $E_6$ gauge
symmetry.

The large DM-nucleon recoil cross section can be readily achieved in the case of a light mediator $Z'$.
Denoting the Dirac or complex scalar DM $U(1)_X$ charge as $Q_{\rm DM}$,
then in light of Eq.~(\ref{fpV}) and Eq.~(\ref{solution}) we have
\begin{align}\label{ISVU}
f_p&=\f{g_X^2Q_{\rm DM}Q_{H_u}}{M_{Z'}^2}\f{2}{1-{\cal I}}\cr
&\approx1.2\times10^{-5}\times \L\f{1\,\rm GeV}{M_{Z'}}\R^2\L\f{g_X^2Q_{\rm DM}Q_{H_u}}{10^{-5}}\R,
\end{align}
where ${\cal I}=-0.7$ has been fixed. The smallness of the product in the second bracket square
can be simply due to a very weak coupling $g_X\ll1$, which is consistent with the lightness of
the dark gauge boson $Z'$. And we will verify this point in the concrete model constructed in
Section~\ref{model:seesaw}.

As mentioned in the introduction, the inelastic DM (iDM)~\cite{iDM} is of special interest. In the iDM scenario,
DM has an exciting state, with mass splitting tens of keV. When DM scatters with nucleons, it dominantly scatter into
its exciting state. The vector interaction mediated by a vector boson $Z'$ can realize the iDM scenario, and in this
case we only need to consider the following interactions
\begin{align}
g_{ab}\bar\chi_a\gamma^\mu \chi_b Z'_\mu,\quad g_{ab}\phi^\dagger_a\tensor\partial^\mu \phi_b Z'_\mu,
\end{align}
where $\chi$ is a pseudo Dirac fermion and $\phi$ is an approximately complex scalar.
The gauge interactions have to be off-diagonal and thus $g_{ab}\propto 1-\delta_{ab}$ at the leading order.
Therefore, if the ISV-iDM is favored, $U(1)_X $ models should receive special attention and
we shall consider them in details later.

\subsection{(Approximate) Spectator Higgs Doublet}\label{IHiggs}

The conventional models, such as the minimal supersymmetric standard model (MSSM), with one or two Higgs doublets as
mediators fail in accommodating the desired ISV. In these models the Higgs doublets not only dominantly
mediate interactions but also account for the fermion masses, and consequently we have the relations
  \begin{align}\label{ratio}
\f{a_{u_i/d_i}}{m_{u_i/d_i}}\propto \f{1}{v_{u/d}},
\end{align}
with $v_{u/d}$ the vacuum expectation value (VEV) of Higgs field $H_{u/d}$, respectively. Such ratios are independent on flavors,
and then the second and third families give the main contribution to $\sigma_p$ due to their larger form factors.
As a result, in such models the DM-nucleon interactions do not show significant ISV.

However, if the Higgs doublet is an (at least approximate) spectator to the electroweak symmetry
breaking (EWSB), ISV may arise. The VEV of a spectator Higgs doublet is zero or very small, and hence
its Yukawa couplings are free parameters. In this way we avoid Eq.~(\ref{ratio}) and
the light quark contributions to $\sigma_p$ can be dominant to allow for ISV.
With such a Higgs, ISV readily arises from the interactions between the spectator Higgs and quarks
\begin{align} \label{Yukawa1}
-\mathcal{L}_Y&\supset \L y_{u,1}\bar q_L \epsilon H_1^\dagger u_R+y_{d,1}\bar q_L H_1 d_R\R+\L
y_{u,2}\bar q_L \epsilon H_2^\dagger u_R+y_{d,2} \bar q_L H_2 d_R\R+h.c.~.~\,
\end{align}
Here $H_1$ is the SM Higgs field breaking electroweak symmetry via its neutral component's
VEV $\langle H_1^0\rangle\equiv v$ around $174~{\rm GeV}$. And $H_2=\L H^+,\frac{H^{0}+i A^0}{\sqrt{2}}\R^T$,
which carries the same SM quantum numbers as $H_1$, is the spectator Higgs with negligible VEV. Thus
Yukawa matrices $y_{u,2}$ and $y_{d,2}$ are free except for constraints from flavor violations.

Now we consider the interactions between the spectator Higgs field and dark sector.
As the $Z'$ mediator case, we do not need to specify a dark matter model. Written
in components, the generic interactions between $H_2$ and the dark sector are given by
\begin{align} \label{SHMD}
&a|\phi|^2H^0,\quad \bar\chi(\alpha-\beta\gamma^5) \chi H^0,\\
&a_A|\phi|^2A^0,\quad \bar\chi(\alpha_A-\beta_A\gamma^5) \chi A^0.
\end{align}
The CP-even component $H^0$ leads to the DM-nucleon SI scattering. Although the CP-odd component
$A^0$ does not contribute to such a scattering, it opens other annihilation channels for DM.
Letters $a,\alpha,$ etc., denote the DM-Higgs effective couplings and they are model dependent.
If DM is a real or complex scalar coming from the Higgs-port models~\cite{Higgspport}, we can readily get
the first term in Eq.~(\ref{SHMD}) via the following term
\begin{align} \label{}
\ld_{\phi}|\phi|^2H_1H_2+h.c.\Rightarrow a=\sqrt{2}\ld_\phi v.
\end{align}
If DM is a SM singlet fermion, the renormalizable interaction between DM and $H^0$ needs extra particles,
e.g., a SM singlet scalar $S$ which further mixes with $H^0$ after EWSB.
If the dark sector contains a term $\ld S\bar\chi\chi$, we then realize the operators given in Eq.~(\ref{SHMD}).
A case in point is the singlet-like neutralino DM within the next to the MSSM~\cite{LDMINMSSM}.

ISV is determined by the Yukawa flavor structure involving $H_2$. Transforming the quarks into the
mass eigenstates, we find that Eq.~(\ref{Yukawa1}) becomes
\begin{align}\label{IHD}
\mathcal{L}_Y\supset &\f{H^{0}-i A^0}{\sqrt 2}\bar u Y_{uu} P_Ru- H^- \bar d Y_{du} P_R u  \cr
                   &+\f{H^{0}+i A^0}{\sqrt 2}\bar d Y_{dd}P_R d+H^+ \bar u Y_{ud}P_R d+h.c.,
\end{align}
where the effective Yukawa coupling matrices are defined by $Y_{qq'}\equiv V_{q_L}^\dagger y_{q',2}V_{q',R}$.
Generally speaking, $Y-$matrices are not diagonal, and consequently the spectator neutral Higgs gives rise to
tree-level flavor changing neutral currents. But $y_{q,2}$ and $V_{q_R}$ are free matrices, thus
the flavor problem in principle can be avoided by properly arranging them (We do not enter into the details here.).
Integrating out $H^0$ we get the effective DM-quark operator coefficients:
\begin{align}\label{IHY}
a_q=&\f{a}{\sqrt{2}(2M_{\rm DM})}\f{1}{m_{H^0}^2}\L Y_{qq}\R_{11},\quad (\rm scalar~DM),\cr
a_q=&\f{\alpha}{\sqrt{2}m_{H^0}^2}\L Y_{qq}\R_{11},\quad (\rm fermionic~DM).
\end{align}
In terms of Eq.~(\ref{ISVSquark}), ISV ${\cal I}\approx-0.7$ is obtained given the ratio
${(Y_{uu})_{11}}/{(Y_{dd})_{11}}\approx -0.77$. Of course,
$(Y_{qq})_{22,33}$ should be sufficiently small so that DM dominantly interacts with the first-family quarks.

Interestingly, recently a spectator Higgs doublet is also introduced to explain the
CDF W+2jets anomaly~\cite{Cao:2011yt,NOVEV}. We now investigate whether or not the spectator Higgs
producing proper ISV can additionally account for this anomaly, where the relevant process is
$p\bar p\ra H^\pm\ra W^\pm H^0(A^0)\ra \ell^+\ell^-\nu +jj$~\cite{Cao:2011yt} and
the measured invariant mass of two jets fixes the neutral Higgs mass, e.g., $m_{H^0}\simeq150$ GeV.
Ref.~\cite{Cao:2011yt} presents a benchmark point to explain the data: $(Y_{uu})_{11}\simeq0.06$ and $m_{H^\pm}\simeq 250$ GeV.
Such parameters in turn means that to obtain $\sigma_p\sim 0.01$ pb we need
\begin{align}\label{}
\rm Scalar~DM:&\quad a\, \L Y_{uu}\R_{11}\sim 1.6{\,\rm GeV},\quad a \,\L Y_{dd}\R_{11}\sim -2.0{\,\rm GeV},\cr
\rm Fermionic~DM:&\quad\alpha\, \L Y_{uu}\R_{11}\sim 0.10,\quad\,\,\quad \alpha \,\L Y_{dd}\R_{11}\sim -0.13~.~
\end{align}
For a real scalar or Majorana DM these values should be reduced by half. This data alone is not problematic,
but there is a potential problem which renders the above analysis invalid.

The problem is that $H^0$ can decay to a pair of LDM,
and its branch-width $\Gamma(H^0\ra 2{\rm DM})$ may exceed $\Gamma(H^0\ra u\bar u)$ which, however, has been
assumed to be the largest one in Ref.~\cite{Cao:2011yt}. Explicitly, their ratios are
\begin{align}\label{}
\f{\Gamma(H^0\ra 2{\rm DM})}{\Gamma(H^0\ra u\bar u)}\sim \f{(Y_{uu})_{11}^{-4}}{3}\L\f{a(Y_{uu})_{11}}{M_{H_0}}\R^2,\quad \f{1}{3}\L\f{\alpha(Y_{uu})_{11}}{(Y_{uu})_{11}^2}\R^2.
\end{align}
for scalar and fermionic DM, respectively. In the above estimation we only consider the decay
$H^0\ra \bar uu$ at the leading order, ignoring QCD corrections. With the aforementioned parameter setting,
we find that for the (real) scalar DM the two channels are comparable, and thus the previous analysis is not affected
substantially. But for the fermionic DM $H^0$ dominantly decays to the DM pair and thus that analysis is no longer valid.
The situation can be improved by considering the alternative process
$p\bar p\ra H^0\ra W^\pm H^\mp\ra \ell^+\ell^-\nu +jj$~\cite{Cao:2011yt} and further inverting the
order of $H^0$ and $H^\pm$ masses. Now $m_{H^\pm}=150$ GeV, so the decay width of $H^\pm\ra W^\pm$+DM+DM is
suppressed by an additional phase factor factor $1/(2\pi)^3$. Moderately increasing $(Y_{uu})_{11}$ and/or lowering
$m_{H^0}$ so as to increase the production rate of $H^\pm$, the fermionic DM may also be consistent with $W+jj$.

Finally we discuss the condition that such a spectator Higgs doublet can appear.
Equivalently we need find when $v_2$ is so small that Yukawa couplings $y_{u/d,2}$ become free parameters.
Roughly speaking, this requires $(Y_{uu})_{11}v_2<m_u,\,(Y_{dd})_{11}v_2<m_d$, where $(Y_{uu})_{11}\sim 0.06$ is
taken to explain $W$+2jets. So we get the condition $v_2<0.05$ GeV.
Now let us  consider the renomalizable scalar potential of the two Higgs doublet model
\begin{align} \label{potential}
V=&\mu_1^2|H_1|^2+\mu_2^2|H_2|^2+\mu_{12}^2(H_1^\dagger H_2+ H_2^\dagger H_1)+\f{\ld_1}{2}|H_1|^4+\f{\ld_2}{2}|H_2|^4 \cr
&+\ld_3|H_1|^2|H_2|^2+\ld_4|H_1^\dagger H_2|^2+\f{\ld_5}{2}(H_1^\dagger H_2)^2+\f{\ld_5^*}{2}(H_2^\dagger H_1)^2,
\end{align}
where all parameters are assumed to be real for simplicity. A small $v_2$ requires an accordingly small Higgs mixing
$\mu_{12}^2$, which can be seen from
the tadpole equation
\begin{align}\label{tadpole}
\f{\partial V}{\partial H_0}=\mu_{12}^2v+\ld_2 v_2^3+ \ld_3v^2 v_2+\ld_4 v^2 v_2+\ld_5 v^2 v_2 =0.
\end{align}
Thus for $\mu_{12}^2\lesssim{\cal O} (10)$ GeV$^2$ we get
$v_2\sim-{\mu_{12}^2 v}/{\L(\ld_3+\ld_4+\ld_5)v^2+\mu_2^2\R}\sim {\cal O}(10)$ MeV with $\ld_{3,4,5}\sim {\cal O}(0.1)$.
In conclusion, it is not difficult to get an approximate spectator Higgs doublet to account for ISV
and the W+$jj$ anomaly simultaneously, at least for scalar DM.

\subsection{Color Triplets}\label{ColorTriplets}

Color triplet, which mediate DM-quark interaction in the $s-$channel and
generate scalar and vector interactions simultaneously, are distinguished from $Z'$
and Higgs mediators. We focus on the fermionic DM
and scalar color triplet mediators case. For the scalar DM and fermionic color triplet mediators case
the discussions are similar and thus not presented.

We allow each quark type $q_{L/R}$ to have its corresponding color triplet mediator $\wt q_{L/R}$,
just like the case in the MSSM. However, only the first family, which is relevant to ISV,  will be considered.
The relevant terms take a general form as (We adopt a MSSM notation):
 \begin{align}\label{TUV}
{\cal L}=&-m_\chi\bar\chi\chi
-\sum_{\alpha,\beta=L/R} m_{u_{\alpha\beta}}^2\wt u_\alpha\wt u_{\beta}^\dagger-\sum m_{d_{\alpha\beta}}^2\wt d_\alpha\wt d_{\beta}^\dagger \cr
&-\sum_{q=u,d}^\alpha\left[\ld_{q_\alpha}\bar \chi (1+\gamma_5)q \wt q_\alpha^\dagger+\ld'_{q_\alpha}\bar \chi (1-\gamma_5)q \wt q_\alpha^\dagger+h.c.\right],
\end{align}
where DM and mediators are assumed to be odd under a $Z_2$ symmetry.
$\wt q_L$ and $\wt q_R$ carry identical $SU(3)_C\times U(1)_{\rm EM}$ quantum numbers and they can violate the
chiral symmetry in two ways, via the direct left-right (L-R) mixing or via simultaneously coupling to $q_L$ and $q_R$ (L-R coupling),
i.e., $\ld_{q_L}\,\ld_{q_R}'\neq0$. The scalar triplets mass eigenstates are denoted by $\wt q_{{1,2}}$, with
corresponding mass eigenvalues $m_{\wt q_{1,2}}$. And $\wt q_{{1,2}}$ are related to the gauge eigenstates by
$\wt q_{\alpha}=\sum_lF_{{q_\alpha}l}\wt q_{l}$ with
 \begin{align}
 F_{q_{L}1}=&\cos\theta_q,\quad F_{q_{L}2}=\sin\theta_q,\quad
  F_{q_{R}1}=-\sin\theta_q,\quad F_{q_{R}2}=\cos\theta_q.\\
  \tan\theta_q=&x_q-\sqrt{1+x_q^2}<0,\quad x_q\equiv \L m_{\wt q_L}^2-m_{\wt q_R}^2\R/2m_{\wt q_{LR}}^2.\label{mixing}
\end{align}
Then the interactions can be rewritten in a from
 \begin{align}
{\cal L}\supset&-
\bar\chi\L\alpha_{l}^q+\beta_{l}^q\gamma_5\R q\wt q_{l}+h.c.,\cr
\alpha_{l}^q=&\sum_\alpha\L\ld_{q_\alpha}+\ld'_{q_\alpha}\R F_{q_\alpha l},\quad \beta_{l}^q=\sum_\alpha\L\ld_{q_\alpha}-\ld'_{q_\alpha}\R F_{q_\alpha l}.
\end{align}

In light of  Eq.~(\ref{STUV}) and Eq.~(\ref{STO}), integrating out $\Phi$ leads to the
effective operators involving SI scattering in the form of Eq.~(\ref{DDS})
and Eq.~(\ref{DDV}), and the operator coefficients are
 \begin{align}
a_{u}=&-\f{1}{m_{\wt u_l}^2}{\rm Re}\L\ld_{u_\alpha}^*\ld_{u_\beta}'F_{u_\alpha l}^*F_{u_\beta l}\R,\cr
b_{u}=&-\f{1}{2m_{\wt u_l}^2}\L\ld_{u_\alpha}\ld_{u_\beta}^*+\ld_{u_\alpha}'\ld_{u_\beta}'^*\R F_{u_\alpha l}F_{u_\beta l}^*~.~
\end{align}
The expressions for $a_d$ and $b_d$ are obtained by replacing $u$ with $d$ in the above equations.
There are two interesting limits. One is the chiral limit $a_{u/d}\ra 0$ which arises when both L-R mixing and L-R
coupling are negligible. In this limit the scalar interactions vanish while vector interactions
leave~\footnote{This fact implies a possibility: We do not have a gauge boson mediator, but a color triplet
mediator in the chiral limit also realizes iDM.}. But this limit fails to produce a negative ${\cal I}$, because
$b_u$ and $b_d$ take the same sign. Therefore this limit should be avoided.
In the other limit, oppositely, only scalar interactions leave, which
happens when DM is a Majorana fermion or real scalar. For general case both scalar and vector interactions do exist and we have
$f_n=a_n+b_n$.

L-R couplings usually are small. As an example, we consider the MSSM with neutralino $\chi_1$ as the
lightest supersymmetric particle (LSP). Its Higgsino component gets L-R coupling from the Yukawa couplings
which however are small. So we take $\ld_{q_L}=\ld_{q_R}'=0$ and consider
L-R mixing as the unique chiral symmetry breaking source, then
 \begin{align}
a_{q}=-\f{\ld_{q_L}'\ld_{q_R}}{m_{\wt q_1}^2}\times\L \f{m_{\wt q_2}^2-m_{\wt q_1}^2}{2m_{\wt q_2}^2}\sin2\theta_q\R\equiv -\f{\ld_{q_L}'\ld_{q_R}}{m_{\wt q_1}^2}{F_q},
\end{align}
with $q=u,d$. We need $F_q\sim{\cal O}(1)$. But if $\wt q_{1,2}$ are degenerate or the L-R mixing angle $\theta_q$ is small,
$F_q$ and hence $a_q$ will be suppressed. According to Eq.~(\ref{ISVSfp}), ${\cal I}=-0.7$, we get
 \begin{align}\label{fpUT}
f_{p}=0.5\times10^{-5}\times\L\f{\ld_{d_L}'\ld_{d_R}}{1}\R\L\f{500\,\rm GeV}{m_{\wt q_1}}\R^2\L\f{F_q}{1}\R\rm GeV^{-2}.
\end{align}
This estimated value should be doubled for complex scalar DM.

We apply the above result to the supersymmetric models with a gauge group $U(1)_X$, and the
light $U(1)_X-$gaugino is the LSP. Moreover, the first-family squarks are the dominant mediators (The second and third
family squarks are negligible due to either their heaviness or their smallness of $U(1)_X$ charges).
Then the relevant terms are
 \begin{align}
{\cal L}\supset&-
\f{\sqrt{2}}{2}Q_{q_L}g_X\wt B_X\L1-\gamma_5\R q\wt q_{L}^\dagger\cr
&-\f{\sqrt{2}}{2}Q_{u_R}g_X\wt B_X\L1+\gamma_5\R u\wt u_{R}^\dagger-\f{\sqrt{2}}{2}Q_{d_R}g_X\wt B_X\L1+\gamma_5\R d\wt d_{R}^\dagger+h.c.,
\end{align}
where $Q$ denotes  $U(1)_X$ charge. The couplings defined in Eq.~(\ref{TUV}) inherit the $U(1)_X$ gauge couplings, and we can extract them from the
above equation:
 \begin{align}\label{YU}
\ld_{d_L}'=&\ld_{u_L}'=\f{\sqrt{2}}{2}g_XQ_{q_L},\quad
\ld_{d_R}=\f{\sqrt{2}}{2}g_XQ_{d_R},\quad
\ld_{u_R}=\f{\sqrt{2}}{2}g_XQ_{u_R}.
\end{align}
But the induced $a_q$ are only semi-quantized since they depend on extra parameters $F_q/m_{\wt q_1}^2$. In the
light of Eq.~(\ref{ISVSquark}) we have
\begin{align}
\f{a_u}{a_d}=\f{Q_{u_R}}{Q_{d_R}}\L\f{m_{\wt d_1}^2F_u}{m_{\wt u_1}^2F_d}\R\simeq-0.77.
\end{align}
Using Eq.~(\ref{mixing}) one can see that $Q_{u_R}\,Q_{d_R}<0$ is necessary to get ${\cal I}<0$, and therefore we
require a chiral $U(1)_X$. It is tempting to regard $U(1)_X$ as $U(1)_Y$ and the LSP as bino,
but quantitatively this possibility is excluded owing to the fact that $a_{u,d}$ are suppressed by large squark mass squares and
small $U(1)_Y$ gauge couplings. The exotic $U(1)_X$ with a relatively large gauge coupling may work.

To end up this subsection we would like to comment on another interesting aspect of the SM extended with color
triplets. In the model Eq.~(\ref{TUV}), the Yukawa couplings $\ld_{q/u/d}$ generically are complex (Namely they
introduce physical CP phases), as opens the possibility to generate sufficient matter asymmetry, which is short within the SM,
via triplets decay. Additionally, DM may also be asymmetric so as to explain
the coincidence $\Omega_{\rm DM}h^2:\Omega_{b}h^2\approx 5:1$ and in turn a light DM around 8 GeV~\cite{NEW}.

\subsection{Dual Mediators}

In the previous discussions we concentrate on ISV coming from a single-type mediator,
and find that no conventional mediator succeeds in giving proper ISV. Dual mediators
bring difference. We have shown that $Z-$boson only mediates the DM-neutron interaction.
But if it interferes with an ordinary Higgs mediator, proper ISV may be produced.
The interactions of the complex sneutrino DM $\wt\nu_1$~\cite{Belanger:2010cd,Kang:2011wb}
fit this scenario. Nevertheless, to guarantee the $Z-$boson invisible decay width (to a pair of DMs)
below the experimentally allowed level, we get $g_{Z_{11}}<0.023$~\cite{Nakamura:2010zzi} with
$g_{Z_{11}}$ the effective coupling constant of the vertex $Z^\mu\wt \nu_1\tensor\partial_\mu \wt\nu_1^*$.
Then we obtain $f_n\approx -M_Z^{-2}g\, g_{Z_{11}}/4\cos\theta_w\sim-0.5\times
 10^{-6}\rm \,GeV^{-2}$, and thus the resulted $\sigma_p$ is much smaller than the desired value.

We consider replacing $Z$ with $X$, a GeV-scale light gauge boson of $U(1)_X$. Here $X$ only interacts
with the SM sector via kinematic mixing with the $U(1)_Y$ gauge boson. As noticed in Ref.~\cite{Kang:2010mh},
this kind of mediator only mediates the DM-proton interaction. To see that, we start from the
gauge kinetic sector of the $U(1)_X\times U(1)_Y$ gauge groups
\begin{align}
{\cal L}_{gauge}&=-{1\over 4}F_Y^{\mu\nu}F_{Y\mu\nu}-{1\over
4}F_X^{\mu\nu}F_{X\mu\nu}+{\theta\over 2}F_Y^{\mu\nu}F_{X\mu\nu}~.~\,
\end{align}
We focus on the small mixing limit $\theta\ll1$, where the $Z$ invisible decay width is not
affected much in the presence of $U(1)_X$ charged LDM. Recall that $X$ mass
$M_X$ should be lighter than DM, then we have $M_{X}<M_{\rm DM}\ll M_Z$, which allows us to
approximate the leading interactions between $X_\mu$ and the SM-sector as~\cite{Baumgart:2009tn}:
\begin{align}
 {\cal L}_{coupling}\supset &\theta X_\mu \L\cos\theta_wJ^\mu_{em}+{\cal
O}(M_X^2/M_Z^2)J_Z^\mu\R,\cr
J_{em}^\mu=&g\sin\theta_w\left[\f{2}{3}\bar u\gamma^\mu u+\L-\f{1}{3}\R\bar d\gamma^\mu d+\L-1\R\bar e \gamma^\mu e\right].
\end{align}
We are working in the mass eigenstate basis $(Z_\mu,A_\mu,X_\mu)$.
The kinematic mixing only induces the coupling between $X_\mu$ and the
electromagnetic current $J^\mu_{em}$. In other words, $X_\mu$ behaves
like a massive photon and only mediates the DM-proton interaction.
As for the DM and $X$ gauge interactions can be written as
\begin{align}
-{\cal L}_{\rm DM}&\supset g_{\rm DM}X_\mu\phi^*\tensor\partial^\mu\phi,\quad g_{\rm DM}X_\mu\bar\chi\gamma^\mu\chi.
\end{align}
With them, we can derive the coefficient
\begin{align}
b_p= \L\f{ g\,\sin2\theta_w}{2}\R\L\f{g_{\rm DM}\theta }{M_{X}^2}\R.
\end{align}

To get a correct ${\cal I}$, we further introduce a Higgs mediator, identified as the SM Higgs $h$.
It is not difficult to get its isospin-conserving contributions to $a_{p,n}$:
\begin{align}\label{dual:ap}
a_p\approx a_n&\approx\f{m_n}{\sqrt{2}(2M_{\rm DM})} \f{a}{v}\f{1}{m_h^2}\L f_{T_s}^{(n)}+3\times\f{2}{27}f_{T_G}^{(n)}\R\cr
&=0.15\times 10^{-5}\L\f{10\,\rm GeV}{M_{\rm DM}}\R\L\f{100\rm\, GeV}{m_h}\R^2\,\rm GeV^{-2}.
\end{align}
A scalar DM $\phi$ is under consideration. In the above equation only contributions from the $s$ quark and heavy quarks are included.
The parameter $a$ is assumed to come from $\ld_\phi|H|^2|\phi|^2$ with $\ld_\phi=1$, but such a large $\ld_\phi$
endangers not only perturbitivity but also naturalness (It renders DM sub-TeV heavy). 2HDM
with a large $\tan\beta$ can alleviate this problem. If $\phi$ couples to the Higgs fields mainly via
$\ld_\phi|\phi|^2H_dH_u$ and moreover the heavier CP-even Higgs boson ($\simeq H_d^0$) is the dominant mediator,
then in Eq.~(\ref{dual:ap}) we get
\begin{align}
\f{a}{v}\ra \f{\sqrt{2}\ld_\phi v_u}{v_d}= \sqrt{2}\ld_\phi\tan\beta.
\end{align}
Now we only need $\ld_\phi\tan\beta/(m_{H_d^0}/100\rm \,GeV)^2\sim1$, which allows $\ld_\phi$ to be
much smaller than the SM-Higgs case. In summary, in the dual mediators case the total DM-nucleon coupling
$f_p=a_p+b_p$ while $f_n=a_n\approx a_p$, and the proper ISV means
\begin{align}
\f{b_p}{a_p}=\f{{\cal I}}{1-{\cal I}}.
\end{align}

Analysis on fermionic DM can be employed similarly, with the replacement $a/2M_{\rm DM}\ra \alpha$ in Eq.~(\ref{dual:ap}).
The numerical problem is exacerbated, and a rather light $m_h$ is required to enhance $a_p=a_n$,
just as the case in the NMSSM with neutralino LSP~\cite{LDMINMSSM}.


 %

\section{A model: Sneutrino iDM with Light $Z'$ Mediator}\label{iDMM}

Ref~\cite{Frandsen} points out that light iDM models with ISV can explain the annual modulation
the CoGeNT and DAMA in a consistent way. As a realization, in this section we propose the sneutrino iDM model with a
very light and very weakly coupled $Z'$ mediator. This model is also an UV completion example of the previous
general analysis.

Before the proceeding discussion on the sneutrino iDM model, we would like to emphasize that
$Z'$ below GeV scale (It is even much lighter than the $Z'$ discussed in~\cite{heavyU(1)})
is of particular interest. In addition to evade the collider bounds, such a light $Z'$ provides
the kinetic block mechanism like in~\cite{Ruderman:2009tj,Kang:2010mh}, which forces DM to mainly annihilate into $2e$ or/and $2\mu$ today,
to avoid the stringent constraints from Solar neutrino, CMB as well as PAMELA.

We explicitly show how does this mechanism work for a scalar iDM. At the early Universe, DM
has two kinds of comparable annihilation channels, DM+DM$\ra q\bar q$ via $Z'$ mediation and
the invisible modes DM+DM$\ra XX$ via contact interactions. $X$ denotes the sub-GeV hidden state
such as $Z'$ which subsequently decays into the safe states $e/\mu/\gamma$. The total
annihilation rate can be parameterized as
\begin{align}\label{ANNI}
\sigma_{an}|v|\sim&\sum_{\pm}\f{\f{1}{4}g_X ^4Q_{\rm DM}^2\L Q_{f_L}\pm
Q_{f_R}\R^2}{16\pi}\f{c_f}{ M_{\rm DM}^2}v^2+\L
\f{g_X ^4Q_{\rm DM}^4}{16\pi}\f{1}{ M_{\rm DM}^2}+...\R,
\end{align}
with $Q_f$ the $U(1)_X$ charge of $f$. The second term collets the invisible modes and today it becomes
the dominant term for $Q_f\sim Q_{\rm DM}$. So the DM annihilation produces no dangerous final states.
But for a Dirac iDM, the modes DM+DM$\ra f\bar f$ have no $v^2$ suppressing and thus the
mechanism fails. But lowering $Z'$ mass even orders lighter or assuming that only $Q_u$ and $Q_d$
are nonzero among $Q_f$ may help to avoid the Solar neutrino constraint.

\subsection{Light Sneutrino iDM from Low Scale Seesaw Mechanism}\label{model:seesaw}

In the MSSM extended with low-scale seesaw mechanism, sneutrino is a natural
scalar iDM candidate (A simple non-supersymmetric iDM model is presented in
Appendix~\ref{NonSUSY}.). To realize ISV as in Section~\ref{ISVU(1)},
the MSSM gauge groups are extended by $U(1)_X$,
and the right-handed neutrino (RHN) is charged under it. The origin of the RHN Majorana mass scale
is attributed to the $U(1)_X$ breaking scale. On our purpose, the relevant terms are
\begin{align} \label{Yukawa2}
W\supset& y_{ij}^N L_i  N_jH_u+\f{\ld_i}{2}S N^2_i+\mu H_uH_d,\\
-{\cal L}_{soft}\supset&\L m_{\wt N_i}|\wt N_i|^2+m_S^2|S|^2\R
+A_0\L y_{ij}^N {\wt L}_i  {\wt N}_jH_u+\f{\ld_i}{2}S {\wt N}^2_i+h.c.\R. \label{YukawaSoft}
\end{align}
All parameters are assumed to be real and only one family of RHN is introduced. The singlet soft mass scale
typically is below one GeV and $A_0$, which controls the mass splitting of sneutrinos, is very small.
Such a soft parameter pattern may be expected if it originates in the gauge mediated SUSY-breaking.

$S$ carries $U(1)_X$ charge $Q_S=-Q_N/2$, and we will show $S$ can spontaneously break $U(1)_X$ at a low scale
$\langle S\rangle\equiv v_s\sim {\cal O}(100)$ GeV. The scalar potential involving $S$ is
\begin{align} \label{}
V_S=&V_F +V_D+V_{soft},\cr
V_F =&|\ld_i SN_i+y_{ji}^NL_jH_u|^2,\cr
V_D=&\f{g_X ^2}{2}\L Q_{H_u}|H_u|^2+ Q_{H_d}|H_d|^2+Q_{S}|S|^2+\sum_f Q_f|\wt f|^2\R^2,
\end{align}
with $V_{soft}=-{\cal L}_{soft}$ given by Eq.~(\ref{YukawaSoft}). In $V_S$, the part relevant
to $U(1)_X$ breaking can be casted into a $\phi^4-$model:
\begin{align} \label{}
V_S\supset& -\mu_S^2|S|^2+\f{\kappa}{4}|S|^4,\cr
\mu_S^2=&\L g_X ^2Q_{H_u}Q_S\cos2\beta\R v^2-m_S^2,\quad \kappa=2{g_X ^2Q_S^2},
\end{align}
where the relation $Q_{H_d}=-Q_{H_u}$ has been used. It is seen that $V_D$ alone, i.e., in the limit $m_S^2\ra0$,
is adequate to trigger $U(1)_X$ spontaneously breaking given $Q_{H_u}Q_S>0$. In this case the solution is
\begin{align} \label{VS}
v_s=\sqrt{\f{2}{\kappa}}\mu_S\approx\sqrt{\cos2\beta}({Q_{H_u}}/{Q_S})^{1/2}\times v,
\end{align}
which is at the weak scale and we will use this result in the following.

We now examine the mass spectrum. In the first, ignoring the small $Z-Z'$ mixing effect,
at leading order the $Z'$ mass is
\begin{align}\label{}
M_{Z'}\approx& \sqrt{2}g_X |Q_S|v_s\simeq \sqrt{2{|Q_{H_u}|}{|Q_S|}\cos2\beta}\times g_X  v~.~\,
\end{align}
Thus $M_{Z'}$ can be naturally at the GeV scale given $g_X \sim 10^{-2}$.
Note that the CP-even component of $S$, denoted as $h_s$, approximately degenerates with $Z'$.
Assuming that the dominant DM annihilation mode is DM+DM$\ra 2h_s$ via the contact interactions from
$|\ld|^2|\wt N||S|^2$, the proper DM relic density requires $\ld\sim0.1$.
Next, the neutralino system
consists of the singlino $\wt S$ and $U(1)_X$ gaugino ${\wt \ld}_X$. For the model under consideration
they have a Dirac mass term $M_{Z'}\wt S{\wt \ld}_X$ and a gaugino mass term
$\f{1}{2}M_{{\wt\ld}_X}{\wt \ld}_X{\wt \ld}_X$, so the smaller eigenvalue is
below $M_{Z'}\lesssim 1$ GeV. In turn, the LSP is the neutralino rather than the 8 GeV sneutrino.
As a solution, we introduce extra singlets $S'$ (They may be necessary to cancel the $U(1)_X$ gauge anomalies)
having terms $M_iSS_i$ or $S^2S_i$ so as to lift the neutralino masses above 8 GeV.
We finally discuss the sneutrino LSP, which is dominated by the RHN sparticle $\wt N$. $A_0$ splits its
CP-even and CP-odd states to form an iDM with mass and splitting
\begin{align}\label{}
M_{\rm DM}^2=m_{\wt N_1}^2=\ld^2 v_s^2+m_{\wt N}^2,\quad \delta\simeq \f{\ld   v_s}{m_{\wt N_1}}A_0.
\end{align}
The RHN Majorana mass should be $M_{N}=\ld v_s\sim m_{\wt N_1}\sim 10$ GeV.
Therefore to get $\delta\sim10^{-5}$ GeV we need an unnaturally small $A_0\sim\delta$.
This problem can be overcome in the model with inverse seesaw mechanism~\cite{Inverseseesaw}, where
the splitting is naturally small because it is suppressed by the light neutrino mass~\cite{Kang:2011wb}.

\section{Discussion and Conclusion}\label{CD}

The light dark matter models with ISV $f_n/f_p\approx-0.70$ and a large DM-nucleon spin-independent cross section
$\sigma_n\sim{\cal O} (0.01)$ pb may provide a way to understand the confusing direct detection
experimental results. Combining with the stringent astrophysical and collider
constraints, we can further deduce the DM properties. In this work, we investigated the possible origin of ISV based on
effective analysis, and found that ISV must arise from the DM and first-family quarks couplings.
To further explore their UV origins, we considered the operators as a result of integrating out the
following mediators:
 \begin{description}
   \item[$Z'$ from $U(1)_X$] The $U(1)_X$ must be chiral,
 and the light $Z'$ is strongly favored.
   \item[Spectator Higgs doublets] Conventional Higgs doublet mediates interactions 
preserving isospin, so we introduce (approximate) spectator Higgs 
doublet whose couplings to the SM quarks are free parameters. Such a Higgs 
 doublet can be additionally used to explain Tevatron CDF $W+jj$ anomaly. 
   \item[Color triplets]  Combining the squarks in the MSSM with 
 exotic $U(1)_X$ (quarks strongly charged under it), we found that the light
 $\wt B_X$ LSP can generate proper ISV via the first-family squark mediation.
 This  scenario is economic furthermore suffers no flavor problem.
       \item[Exotic $Z'$ plus Higgs] For a SM-neutral $U(1)_X$ having 
kinetic mixing with $U(1)_Y$, its   light gauge boson $X_\mu$ only 
mediates DM-proton interaction. Combining it with a conventional Higgs
 mediator, we can obtain the desired ISV. Similar point is also adopted in
 Ref.~\cite{DelNobile:2011je} during the completion of this work.
 \end{description}
As a realistic model building, we propose the MSSM with low scale seesaw mechanism and sub-GeV scale
$U(1)_X$ gauge group extension. In this model a light sneutrino plays the role of isospin-violating-iDM to
explain the CoGeNT annual modulation, in consistent with other detections results and various bounds.

.

\section{Acknowledgments}

We would like to thank Yang Bai, Wanlei Guo and D. Sanford for useful discussion.
This research was supported in part
by the DOE grant DE-FG03-95-Er-40917 (TL), and
by the Natural Science Foundation of China
under grant numbers 10821504 and 11075194 (TL).

\appendix

\section{A Non-Supersymmetric Scalar iDM Model}\label{NonSUSY}

  In the non-supersymmetric case, there is a simple way
to realize  iDM at renormalizable level. The dark sector consists of two SM singlets $\phi_{1,2}$ with  mass hierarchy ${\cal O}(8{\,\rm GeV})^2\sim m_{\phi_1}^2\ll m_{\phi_2}^2$.
 The relevant   scalar potential for iDM generation is quite simple:
 \begin{align}\label{U1B}
-V\supset
&\L m_{\phi_1}^2|\phi_1|^2+m_{\phi_2}^2|\phi_2|^2\R+\L\eta_1\phi_1\phi_2^*S^2+\eta_2\phi_2^2S^2+h.c\R.
\end{align}
We need to arrange  charge assignment properly so that
 the dark sector  conserves the $Z_2$ symmetry to protect DM stable.
Singlet $S$ breaks $U(1)_X$ at $v_s\sim{\cal O}(100)$ GeV as the model given in the text, and this VEV
also induces a quartic mass term for the heavy state $\phi_2$
\begin{align}\label{}
-V\supset\eta_2v_s^2 \phi_2^2+c.c.,
\end{align}
At the same time, the dark global $U(1)$ symmetry which acts on $\phi_i$ only is broken.
This leads to a small mass splitting between the CP-even and CP-odd states of $\phi_1$.
To extract out the splitting analytically, we diagonlize the mass matrix for $\phi_{1,2}$ by
the unitary matrix $U_\phi$, where we have ignored the $U(1)$ breaking mass term. Then we obtain
\begin{align}\label{MassB}
\phi_1\ra \cos\theta_{12}\phi'_1+\sin\theta_{12}\phi'_2,\quad \phi_2\ra -\sin\theta_{12}\phi'_1+\cos\theta_{12}\phi'_2,
\end{align}
 where fields with primes are in the (approximate) mass eigenstates. The mixing angle is
\begin{align}\label{MG}
\theta_{12}\simeq \eta_1\f{v_s^2}{m_{\phi_2}^2}\ll 1,
\end{align}
which is invalid if $(\eta_1 v_s^2)^2<m_1^2m_2^2$ (To assure the positivity)
and $|\eta v_1v_2|,m_1^2\ll m_2^2$. With it, substituting
   Eq.~(\ref{MassB}) into Eq.~(\ref{U1B}), then we get the mass splitting:
\begin{align}\label{MG}
\delta=2\eta_2\sin\theta_{12}^2\times\f{v_s^2}{m_{\phi_1}}\approx 2\eta_2\eta_1^2\L\f{v_s}{m_{\phi_2}}\R^4\f{v_s^2}{m_{\phi_1}}.
\end{align}
$\delta\sim 10^{-5}$ GeV can be obtained in many ways, even setting
$m_2$ at the TeV scale.

\section{Operators after Integrating Out Mediators}\label{CS}

The models on the DM-SM fermions interactions can be classified based on the propagators mediating DM and SM particle interactions.
In this appendix, we borrow some results from the Ref.~\cite{Agrawal:2010fh}.
 For the scalar DM, 
it interacts with SM fermions by exchanging  a $Z'$ boson, a (real) Higgs doublet
 $h$ and the colored fermion $Q$,  the corresponding  Lagrangian is given by
 \begin{align}
   \mathcal{L}
  &=
  -\frac{1}{4}F'^{\mu\nu}F'_{\mu\nu}
  +\frac{1}{2}m_{Z'}^2\;Z'^{\mu}Z'_{\mu}
  +a\phi^{\dagger}\tensor\partial_{\mu}\phi
  Z'^{\mu}
  +\bar{q}\gamma^{\mu}(\alpha-\beta\gamma^5)q\;Z'_{\mu},\\ \mathcal{L}
  &=
  \frac12 (\partial h)^2 - \frac12 m_h^2 h^2
  -a\phi^{\dagger}\phi h
  - \bar{q}(\alpha-\beta\gamma^5)q h,\\
{\cal L}&=\bar Q\L i \partial -m_Q\R Q-\bar q\L\alpha -\beta \gamma^5\R Q\phi^\dagger-h.c..
\end{align}
Integrating out the heavy propagators via equation of motion, we obtain the
 effective operators generating SI cross section (Other operators belongs to
Eq.~(\ref{ID}), we do not list here)
\begin{align}
  {\mathcal L}_{eff}
  &\supset
  -\frac{a\alpha}{m_{Z'}^2}
  (\phi^{\dagger}\tensor\partial_\mu\phi),\;
  \\
   {\mathcal L}_{ eff}  &\supset
  \frac{a\alpha}{m_h^2}
  \phi^{\dagger}\phi\;
  \bar{q}\,q,~,~\\
{\cal L}_{eff}&\supset\f{1}{m_Q}\L|\alpha|^2-|\beta|^2 \R \bar qq \phi^\dagger\phi
+\f{i}{m_Q^2}\L|\alpha|^2+|\beta|^2 \R \bar q\gamma^\mu q\phi^\dagger\tensor\partial_\mu\phi.
\end{align}
For the real scalar DM, vector interactions disappear.  Interactions of fermionic DM  can be described  analogous to scalar DM
\begin{align}
  \mathcal{L}&=
  -\frac{1}{4}\mathcal{F}'^{\mu\nu}\mathcal{F}'_{\mu\nu}
  +\frac{1}{2}m_{Z'}^2\;{Z'}^{\mu}{Z'}_{\mu}
  +\bar{\chi}\gamma^{\mu}
  (\alpha-\beta\gamma^5)\chi {Z'}_{\mu}
  +\bar{q}\gamma^{\mu}(\widetilde\alpha-\widetilde\beta\gamma^5)q {Z'}_{\mu},\\
   \cal{L}
  &=
  \frac12 (\partial h)^2 - \frac12 m_h^2 h^2
  -\bar{\chi}(\alpha-\beta\gamma^5)\chi h
  -\bar{q}(\widetilde\alpha-  \widetilde\beta\gamma^5)q h,\\
    \cal{L}
  &=
  |\partial \Phi|^2-m_\Phi^2|\Phi|^2
  -\bar{\chi}(\alpha-\beta\gamma^5)q \Phi
  -h.c.~,~\label{STUV}
\end{align}
where $\Phi$ denotes the scalar color triplet mediators.
And the corresponding effective operators are
\begin{align}
  {\cal{L}}_{eff}
  &\supset -\frac{1}{m_{Z'}^2}
  \alpha\widetilde\alpha\;
  \bar{\chi}\gamma^{\mu}\chi \;
  \bar{q}\gamma_{\mu}q,
 \\
  {\cal{L}}_{eff}  &\supset
  \frac{\alpha\wt\alpha}{m_h^2}
  \bar{\chi}\chi\;\bar{q}q,\\
    {\cal{L}}_{eff}
  &\supset \f{1}{4m_\Phi^2}\left[\L|\alpha|^2-|\beta|^2\R\bar \chi\chi\bar qq+\L|\alpha|^2+|\beta|^2\R\bar \chi\gamma_\mu\chi\bar q\gamma^\mu q\right]~.\label{STO}
\end{align}
When the DM is a Majorana fermion, the vector interaction vanishes.

\section{Scattering and Annihilating}\label{SAA}

In this appendix we briefly introduce the formula involving  direct detections and give the relevant annihilation rates.
Ignoring  small ISV from the form factor of proton and neutron, the DM-nucleus SI scattering cross section at the zero momentum transfer
(It is not the actual cross section $\sigma_n$) can be written in a form~\cite{Jungman:1995df}
 \begin{align}\label{DNS}
\sigma_0=\f{\delta_C\mu_N^2}{\pi}\left[Zf_p+(A-Z)f_n\right]^2,
\end{align}
where $A$ is the atomic mass of the nucleus while $Z$ is its atomic  number. The reduced mass
$\mu_N=M_{\rm DM}m_N/(M_{\rm DM}+m_N)$, and $\delta_C=4$ for the self-conjugate particle like
Majorana and real scalar DM, otherwise $\delta_C=1$. In references DM-proton scattering cross
section is used frequently and it is defined as
 \begin{align}\label{DnS}
\sigma_p=\f{\delta_C\mu_p^2}{\pi} f_p^2,
\end{align}
where $\mu_p$ is the DM-proton reduced mass. $f_p$ coming from the scalar interaction in Eq.~(\ref{DDS})
and vector interaction in Eq.~(\ref{DDV}) are respectively given by
 \begin{align}\label{}
{\rm Fermionic\,\,DM}:&\quad f_p=\sum_q B_q^pa_q;\quad\quad\quad f_p= 2a_u+a_d,\cr
{\rm Scalar\,\,DM}:&\quad f_p=\sum_q B_q^p\f{a_q}{2M_{\rm DM}};\quad f_p= 2a_u+a_d.
\end{align}

The DM can annihilate into the SM particles, which determines the final DM relic density
after the annihilation freeze-out. And the thermally average annihilation cross section,
expanded with relative velocity $v_{rel}$ (The subscript will be omitted), takes a form of
 \begin{align}\label{}
\langle \sigma v\rangle_{F.O.}=a+ b\langle v^2\rangle=\L a+ 3b/x_f\R
\end{align}
 where $x_f\equiv M_{\rm DM}/T_f=3/\langle v^2\rangle$ with $T_f$ the DM decoupling temperature,
 and $x_f\sim20-30$ for the weakly interactive massive particle.
 The DM relic density can be formulated as
 \begin{align}
\Omega h^2\approx \frac{1.07\times10^9 x_f {\rm GeV^{-1}}}{M_{\rm Pl}\sqrt{g_*}\langle \sigma v\rangle_{F.O.}},
\end{align}
with $g_*$ as the effective relativistic degrees of freedom when DM decouples.
The actual effective annihilation rate, which is used to determine relic density
and calculate DM annihilation signals, is given by
 \begin{align}
\langle \sigma v\rangle_{F.O}=T_{\rm DM}\times \L a+3b/x_f\R~,
\end{align}
where $T_{\rm DM}=1/2$ for the complex DM while $T_{\rm DM}=4$ for
 the self-conjugate DM~\cite{Srednicki}.

 The $a$ and $b$  can be extracted out from  partial wave expansion of the cross section times the relative velocity $\sigma v=a+b v^2$ (regarded as a rough thermal averaged cross section). Here the relative velocity is $v=2\sqrt{1-4M_{\rm DM}^2/s}$ with $s$ the Mandelstam variable.
 For the fermionic DM,  $\sigma v$ from  operators involved given by~\cite{Beltran:2008xg},
\begin{align}
a_f\bar \chi\chi\bar f f&:\quad \quad  \f{c_f}{16\pi}\times2a_{f}^2 M_{\rm DM}^2 {\beta_f ^3 }\,v^2,\\
\f{G_{P,f}}{\sqrt{2}}\bar \chi\gamma^5\chi\bar f\gamma^5 f&:\quad \quad\f{c_f}{4\pi}\times G_{P,f}^2 M_{\rm DM}^2 \beta_f  ,\\
\f{G_{PS,f}}{\sqrt{2}}\bar \chi\gamma^5\chi\bar f f&:\quad \quad\f{c_f}{4\pi}\times G_{PS,f}^2 M_{\rm DM}^2 \beta_f  ^3,\\
\f{G_{SP,f}}{\sqrt{2}}\bar \chi\chi\bar f \gamma^5f&:\quad \quad\f{c_f}{16\pi}\times G_{SP,f}^2 M_{\rm DM}^2 \beta_f ,\\
b_f\bar \chi\gamma_\mu\chi\bar f\gamma^\mu f&:\quad \quad\f{c_f}{4\pi}\times 2b_f^2 M_{\rm DM}^2 \beta_f  \L2+z_f\R,\\
\f{G_{A}}{\sqrt{2}}\bar \chi\gamma^5\gamma_\mu\chi\bar f\gamma^5\gamma^\mu f&:\quad \quad \f{c_f}{4\pi}\times G_{A,f}^2 M_{\rm DM}^2 \beta_f  \left[z_f+\f{1}{12}\L4-z_f\R v^2\right],\\
\f{G_{AV,f}}{\sqrt{2}}\bar \chi\gamma^5\gamma_\mu\chi\bar f\gamma^\mu f&:\quad \quad \f{c_f}{48\pi}\times G_{AV,f}^2 M_{\rm DM}^2  \beta_f^2\, v^2,\\
\f{G_{VA,f}}{\sqrt{2}}\bar \chi \gamma_\mu\chi\bar f\gamma^5\gamma^\mu f&:\quad \quad \f{c_f}{2\pi}\times G_{VA,f}^2 M_{\rm DM}^2 \beta_f ,\\
\f{G_T}{\sqrt{2}}\bar \chi\sigma^{\mu\nu}\chi\bar f\sigma_{\mu\nu} f&:\quad \quad\f{c_f}{4\pi}\times  G_{T,f}^2 M_{\rm DM}^2 \beta_f  \L7+ z_f\R,
\end{align}
where the final state velocity  $\beta_f  \equiv \sqrt{1-z_f}$ and $z_f\equiv m_f^2/M_{\rm DM}^2$, the color factor $c_f=3$ for quarks otherwise 1. The scalar DM and relevant operators
$\sigma v$ are given by
 \begin{align}
a_f|\phi|^2\bar f f&:\quad \quad \f{c_f}{8\pi}\times 2a_f^2  \beta_f^3,\\
\f{F_{Vf}}{\sqrt{2}}\phi^\dagger\tensor{\partial}_\mu\phi\bar f\gamma^\mu f&:\quad \quad \f{c_f}{4\pi}\times F_{Vf}^2 M_{\rm DM}^2 \beta_f  \left[\f{2}{3}\L2+z_f\R v^2\right],\\
\f{F_{SPf}}{\sqrt{2}}|\phi|^2\bar f\gamma_5 f&:\quad \quad\f{c_f}{8\pi}\times  F_{SPf}^2  \beta_f  ,\\
\f{F_{VAf}}{\sqrt{2}}\phi^\dagger\tensor{\partial}_\mu\phi\bar f\gamma^\mu\gamma^5 f&:\quad \quad \f{c_f}{4\pi}\times F_{VAf}^2 M_{\rm DM}^2 \beta_f  \left[\f{2}{3}\L2-z_f\R v^2\right].
\end{align}

\end{document}